\documentclass[aps,final,12pt,tightenlines]{revtex4}%
\usepackage{amsfonts}
\usepackage{amsmath}
\usepackage{amssymb}
\usepackage{graphicx}%
\providecommand{\U}[1]{\protect\rule{.1in}{.1in}}

\begin{document}
\preprint{ }
\title[Short title for running header]{Gravitational Wave Test of the Strong Equivalence Principle}
\author{C. S. Unnikrishnan}
\affiliation{\textit{Gravitation Group, Tata Institute of Fundamental Research, Homi Bhabha
Road, Mumbai - 400 005, India}}
\author{George T. Gillies}
\affiliation{\textit{School of Engineering and Applied Science, University of Virginia},
\textit{Charlottesville, VA 22904-4746, USA}}
\pacs{PACS number}

\begin{abstract}
The Strong Equivalence Principle (SEP) holds the full essence and meaning of
the General Theory of Relativity as the nonlinear relativistic theory of
gravitation. It asserts the universal coupling of gravity to all matter and
its interactions including the gravitational interaction and the gravitational
self energy. We point out that confirming the gravitational coupling to
gravitons, and hence to the gravitational waves, is the direct test of the
SEP. We show that the near simultaneous detection of gravitational waves and
gamma rays from the merger of binary neutron stars provides a unique and the
most precise test of the SEP, better than a part in $10^{9}$, which is also
the only test of the SEP in the radiation sector.

\end{abstract}
\startpage{1}
\endpage{10}
\maketitle
\tableofcontents

\vspace{0.5in}

%\volumeyear{year}
%\volumenumber{number}
%\issuenumber{number}
%\eid{identifier}
%\date[Date text]{date}
%\received[Received text]{date}

%\revised[Revised text]{date}

%\accepted[Accepted text]{date}

%\published[Published text]{date}

\section{Introduction}

The Einstein Equivalence Principle (EEP) that asserts the local equivalence of
all physical phenomena in a uniform gravitational field $g$ and an accelerated
frame with $a=-g$ was postulated by Einstein on the basis of the weak
equivalence principle (WEP), \ or the universality of the ratio of the
gravitational to inertial mass. The WEP in turn rests empirically on the
observed Universality of Free Fall (UFF), which is the fact that the
gravitational acceleration does not depend in the mass or other material
properties of the body in free fall. Due to the equivalence of the mass and
energy, UFF implies that the gravitational acceleration does not depend on the
binding energies of the standard model interactions in the falling body. The
formulation of the General Theory of Relativity as a geometrical theory
presupposes that the EEP includes the gravitational phenomena as well. Then,
one has to insist on the UFF of the gravitational binding energy of the
falling bodies, leading to the Strong Equivalence Principle (SEP).
Generalizing from the equivalence of the gravitational mass and the inertial
mass, The SEP asserts the universal coupling of gravity to all matter and
their interaction energy, including that of the gravitational interaction,
highlighting the nonlinear nature of gravity. Hence, \emph{at the fundamental
microscopic level, the SEP affirms the universal gravitational coupling to
gravitons themselves}. This completes the Einstein Equivalence Principle as
the grand generalization that asserts the local equivalence of all physical
phenomena, including gravitational phenomena, in a uniform gravitational field
and in a uniformly accelerated frame. The SEP is the final frontier of
explicit tests of the principle of equivalence \cite{Will,Unni-Gillies-EP}.

A significant test of the SEP requires that the precision in the measurement
of the differential acceleration $\delta a/g$ in the gravitational field $g$
is better than the fractional amount of of the gravitational self energy
relative to the total energy $\varepsilon_{g}/E$. Since $E$ is essentially the
rest mass energy of a massive test body, $\varepsilon_{g}/Mc^{2}\simeq
GM^{2}/R(Mc^{2})=GM/Rc^{2}$. Hence, a significant test requires the
observational access to the relative trajectories of a pair of planetary to
stellar scale test objects, in the gravitational field of a third body.

The SEP has been tested and verified directly in the `free fall' of the Earth
and the Moon towards the Sun employing the Nordtvedt effect
\cite{Nordtvedt,Will}, by monitoring the orbital distance of the Moon from the
Earth in Lunar Laser Ranging (LLR) \cite{Merkowitz,Hofman}. This feat was
possible because of the impressive long term precision achieved, of several
millimeters, in the determination of the orbital distance of $3.7\times10^{8}$
m. This translated to a test of the UFF and the WEP with a precision of 2
parts in $10^{13}$. With the gravitational self energy of $E_{g}/mc^{2}%
\simeq5\times10^{-10}$ for the Earth, and relatively negligible self energy
for the Moon, one obtains the significant observational constraint\ on any
violation of the SEP of
%TCIMACRO{\TEXTsymbol{<} }%
%BeginExpansion
$<$
%EndExpansion
$0.04\%$, which is remarkable considering the smallness of the contribution of
the gravitational self energy of these bodies to their total mass-energy.

The only better observational constraint on the SEP is obtained by comparison
of the orbits of a triple stellar system PSR J0337+1715, consisting of a
binary system of a millisecond radio pulsar (366 Hz) and a white dwarf, with a
1.6-day orbital period, that itself is in a much longer period orbit (327
days) with another white dwarf \cite{Ransom,Archibald,Voisin}. \ The orbits
can be monitored with the radio pulses from the neutron star (pulsar). The
gravitational self energy of a neutron start is about 10\%, and it is
comparatively negligible for the while dwarf. With the limit of 2 parts in
$10^{6}$ on the universality of free fall in this triple stellar system, the
SEP is established with a precision of $10^{-5}$, which is more than an order
of magnitude tighter than the LLR result. However, the analysis involved in
obtaining this constraint is necessarily more elaborate and complicated, owing
the fact that the entire information on the orbits is extracted from the
observation of the pulses from the pulsar in the system.

There is evidently a paucity of accessible tests of SEP, owing to the extreme
weakness of the gravitational interaction. There is only one Earth-Moon-Sun
system in this universe for which we have easy access to, for the study of
such a deep and characteristic foundational issue of gravity. Pulsar-white
dwarf systems provide test systems to study situations in which there is a
large contribution of the gravitational self energy to the total mass energy,
but the precision is limited because the galactic gravitational field in which
they free fall is very small. An observationally convenient and accessible
triple stellar system is a rare chance that is very hard to come by. In this
context, we have found an entirely new way for the precision tests of the SEP
and GTR, leveraging the realization that the gravitational waves are the
purest and manifest form of gravitational energy. Gravitational waves
conceptually correspond to real gravitons, whereas the notional gravitational
self energy in bulk matter corresponds to the virtual carriers of the
interaction. In fact, the binding energy or the self energy, both in case of
electromagnetism and gravitation, is the absence of real energy, rather than
its presence; that much positive energy is released in making the bound
system. The manifest form of the gravitational energy is the gravitational
wave, just as the manifest form of the electromagnetic energy is the
electromagnetic wave.\ Therefore, testing for \emph{the\ gravitational
coupling of the gravitons themselves is the ultimate test of the SEP}. The
photons can provide the ideal reference for the comparison. \emph{Clearly, a
test of the universal coupling of gravitational waves to the gravity of bulk
matter is the most direct test of SEP}.

A new window for this direct and reliable precision test of SEP opened with
the detection of near simultaneous gravitational waves (GW) and gamma rays
from the merger of binary neutron stars \cite{GW170817,Fermi-LSC}. This unique
test has the clear possibility of scores of future detections, allowing
crucial statistical reliability. \emph{The central idea of the test stems from
the realizations that gravitational waves are propagating form of pure
gravitational energy, released directly from the gravitational binding energy
of the binary system}. Then, the gravitational effects of massive structures,
like galaxy clusters, on the propagation of pristine gravitational energy,
relative to the same effects on photons, is a transparent direct test of SEP.
The gamma rays serve as the reference matter-energy, without any significant
gravitational energy, providing a complete test. The Shapiro delay is a first
order test, proportional to the integrated gravitational potential in the
intervening space \cite{Shapiro}, and the gravitational bending is a second
order test, proportional to the gradient of the potential. Hence, the Shapiro
delay provides a much better test of the SEP than the gravitational bending.
But the latter is the one that resembles the `free fall' tests. We examine
both effects for the completeness of the discussion.

\section{Sensitivity of the tests of the SEP}

The classic test of the SEP along the lines of the tests of the universality
of free fall involves comparing the relative trajectory of two bodies \ in the
gravitational fall towards a third massive body. What is measured is the
difference in acceleration $\delta a$ in the gravitational field $g$. The
ratio $\delta a/g$ is identical to the E\"{o}tv\"{o}s WEP parameter $\eta$.
When the relative contribution of the gravitational self energy is larger than
the sensitivity $\eta$ if the test, one gets a useful test of the SEP. \ There
are three key quantities to consider while estimating the precision of the
test of the SEP, denoted by the symbol $\Delta$. One is the gravitational
acceleration at the location of the falling bodies, $g$. The second parameter
is the difference in the ratio of the gravitational energy in the falling
bodies to their total energy. Since the total energy is essentially the rest
mass energy,
\begin{equation}
\delta\varepsilon_{g}=\left\vert \frac{\varepsilon_{g1}}{m_{1}c^{2}}%
-\frac{\varepsilon_{g2}}{m_{2}c^{2}}\right\vert
\end{equation}
The third parameter is of course the precision achieved in the test of the
universality of free fall from factual observation (including both the
statistical and systematic errors). This is the precision to which the
differential acceleration of free fall is constrained relative to the local
gravitational field, $\delta a/g$. \ Clearly, for a given precision in the
measurement of the differential acceleration, a better test is obtained when
the actual gravitational acceleration $g$ is larger. The precision of the test
of WEP is determined by
\[
\eta\equiv\left(  \frac{m_{i}}{m_{g}}\right)  _{1}-\left(  \frac{m_{i}}{m_{g}%
}\right)  _{2}=\delta a/g
\]
Then the sensitivity for the test of SEP is given by $\Delta=\eta
/\delta\varepsilon_{g}$.

For the solar system tests, $g\simeq10^{-2}-10^{-3}$ m/s$^{2}$ and the
precision (2$\sigma$) achieved in observing $\delta a/g$ (for LLR) is about
$10^{-13}$ m/s$^{2}$. \ But the quantity $\delta\varepsilon_{g}$ for the
Earth-Moon system is only $4.6\times10^{-10}$. Therefore, the constraint on
the violation of the SEP is limited to
\begin{equation}
\Delta=\frac{\eta}{\delta\varepsilon_{g}}\leq\frac{10^{-13}}{4.6\times
10^{-10}}\simeq2\times10^{-4}%
\end{equation}
The test employing the free fall of a neutron star-white dwarf system towards
the galaxy can take advantage of the much higher gravitational energy of the
neutron star ($\delta\varepsilon_{g}>10^{-1}$), but much of that advantage is
offset by the tiny galactic gravitational acceleration of only $g\simeq
2\times10^{-10}$ m/s$^{2}$. Yet, impressive constraints on $\eta$ have been
obtained of the order of $\eta\leq10^{-3}$, which translates to the constraint
on SEP of $\Delta=\eta/\delta\varepsilon_{g}\leq10^{-2}$ \cite{Freire}.

In contrast, a triple stellar system like the PSR J0337+1715 in which a
neutron star-white dwarf system orbits another neutron star, allows the better
constraint through painstaking analysis of the orbital data and careful
modeling \cite{Archibald,Voisin}. \ The gravitational acceleration (with the
327 day orbit) is comparable to the LLR case, $g\simeq5\times10^{-3}$ and the
difference in the gravitational self energy is $\delta\varepsilon_{g}%
\simeq10^{-1}$. The precision achieved in estimating the differential
acceleration is $\eta\leq2\times10^{-6}$. This provides the strongest
constraint on SEP to date of $\Delta\leq2\times10^{-5}$. \ 

The GW test of the SEP is a class apart from the tests involving massive
bodies in two respects. A gravitational wave is pristine gravitational energy,
propagating over cosmological distances. Therefore, the ratio of the
gravitational energy to the total energy is unity for GW, $\varepsilon
_{g}/E=1$. Then we need another entity that can co-propagate with GW with
negligible gravitational energy, for comparison, and that is light, for which
$\varepsilon_{g}/E\simeq0$. Thus $\delta\varepsilon_{g}=1$. \ The differential
comparison is possible if both the GW signal and the electromagnetic signal
from the same source is observed. Fortunately, we have one confirmed
astrophysical event with these criteria satisfied, in the gravitational wave
detection in the LIGO-Virgo detectors and the Gamma ray detection by the Fermi
satellite, GW170817+GRB \cite{GW170817,Fermi-LSC}.

\section{GW tests of the SEP}

The triad of interferometric gravitational wave (GW) detectors, LIGO\_Hanford,
LIGO\_Livingston, and Virgo, sensed the arrival of gravitational waves from
the inspiral and the merger of a binary neutron star (BNS) system, on the 17th
August 2017 \cite{GW170817}. Named GW170817+GRB, it became the one-of-a-kind
event (so far) because of the near simultaneous detection of gamma rays,
within about 1.7 s, by the Fermi gamma ray satellite \cite{Fermi-LSC}. This
detection with the three GW detectors allowed the delineation of a relatively
precise localization area in the sky, which led to the identification of the
source galaxy as NGC 4993 at a distance of 40 megaparsecs ($10^{24}$ m). What
is relevant for our test of the SEP is that the gravitational waves and the
gamma rays have to traverse this vast distance in the gravitational presence
of the mighty Virgo cluster of galaxies, before they pass through the
gravitational field of our Milky Way galaxy, and arrive on the Earth.%

\begin{figure}
[ptb]
\begin{center}
\includegraphics[
height=1.9053in,
width=4.9398in
]%
{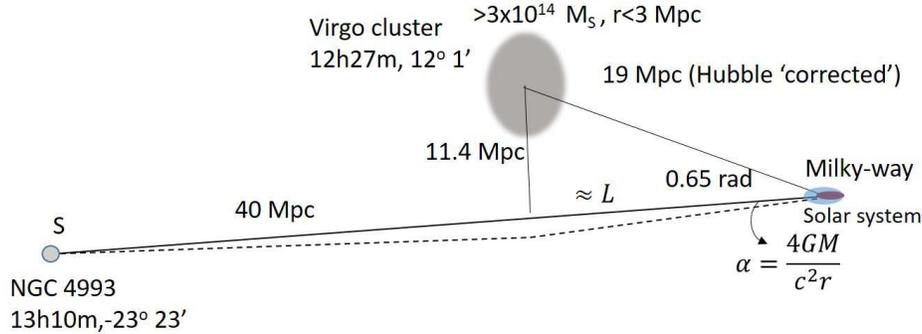}%
\caption{The relative positioning and parameters of the source galaxy, the
solar system and the Virgo galaxy cluster for the calculation of the
propagation delay and gravitational bending.}%
\end{center}
\end{figure}
%EndExpansion

The geometrical configuration of the GW-GRB event relative to the Earth and
the Virgo cluster is indicated in the figure 1. The universal gravitational
coupling of the Virgo cluster mass to the gravitational energy of the GW and
the electromagnetic energy of the gamma rays would manifest in two physical
effects. One is the Shapiro time delay which of first order in $\alpha
=\phi/c^{2}\simeq GM/Rc^{2}$, and the second is the \emph{time delay} due to
the gravitational bending, which is of second order in $\alpha$ (the
gravitational bending itself, which is first order in the potential, cannot be
observed for the gravitational waves).

The Shapiro delay is the excess propagation delay of the waves due to the
gravitational potential, $\int ds(2\phi/c^{3})$. Any violation of the SEP due
to a nonuniversal coupling to the gravitational energy of GW will alter this
delay, relative to the Shapiro delay of the photons. One already know that the
electromagnetic energy in bulk matter obeys the UFF to the very high precision
of $\eta_{em}\leq10^{-9}$, because the ratio of the total electromagnetic
energy in the atom to its rest mass energy is $E_{em}/mc^{2}>10^{-5}$, and the
best tests of the WEP have reached $\eta\leq1.5\times10^{-14}$
\cite{Adel,Microscope}. Therefore, we the gravitational coupling of the gamma
rays would obey the WEP to better than a few parts in $10^{9}$. With the
galactic potential $\phi_{mw}/c^{2}\simeq10^{-6}$, the Shapiro delay during
the propagation over 30 Kpc ($\sim10^{21}$ m) in the Milky Way galaxy itself
amounts to about 100 days, as has been analyzed in a comparison of the
velocity of light and GW in the BNS merger event GW170817+GRB \cite{Fermi-LSC}%
. However, we should consider the Virgo cluster, because the main contribution
of the gravitational potential in our cosmological neighbourhood is from this
galaxy cluster. The Shapiro delay is scalar cumulative effect, steadily
increasing with the same sign. There \ are no other theoretical complication
like fixing a metric because we know with certainty that the propagation is in
the unique $k\simeq0$ FLRW metric of our factual universe. \ Therefore, the
calculation with the mass of the Virgo cluster gives a reliable lower limit to
the gravitational effect between NGC 4993 and the Earth. Because the
propagation duration $t$ (over the distance $>$10 Mpc) in the average potential $\phi_{V}/c^{2}\simeq1.5\times10^{-6}$ of the
Virgo cluster is more than $10^{15}$ s, a very conservative value for the
gravitational Shapiro delay is readily estimated from $\delta t=\int
ds(2\phi/c^{3})$ as
\begin{equation}
\delta t>\frac{2\phi_{V}}{c^{2}}t\simeq3\times10^{9}~s
\end{equation}
which is to be compared with the observed 1.7 s between the arrival times of
the gravitational waves and the gamma rays. This implies the gravitational
coupling of propagating gravitational energy to the source masses is the same
as the gravitational coupling of electromagnetic energy within $\Delta
\simeq6\times10^{-10}$. Note that any kind of electromagnetic factor along the
path can only introduce further delays in the propagation of the photons;
hence the factual constraint is always better than our conservative
constraint. The similar propagation of the GW and the gamma rays from
GW170817+GRB has yielded a very high precision test of the SEP at better than
part in a billion!

We can also get a constraint on the SEP from the similar bending of the GW and
light in the field of the Virgo cluster, albeit with lower precision. This is
similar to the free fall test of two bodies in the gravitational field of a
third body, with the additional aspect of relativistic propagation of the
tested entities. If there are many massive structures, distributed around the
path of propagation, the precise calculation of the resultant gravitational
bending is complicated, and requires the detailed catalogue of matter
distribution. Unlike the scalar Shapiro delay, the gravitational bending by a
large structure like the Virgo cluster can be partially nulled, if there are
several smaller structures that are closer to the path of propagation. If the
gravitational waves and the gravitational energy in them did not experience
the same `free fall' as the photons in the gravitational field of the Virgo
cluster, there would have been significant discrepancy in the time of arrival
of the gravitational waves and the gamma rays.

The angle of bending due to the gravity of a compact source is well known as
$\alpha=4GM/c^{2}R$. Since the bending is small, the difference in the
distance of propagation between the deflected path and the unperturbed path is
(figure 1),
\begin{equation}
\delta L=\frac{L}{\cos\alpha}-L\simeq\frac{L}{1-\alpha^{2}/2}-L\simeq
L\alpha^{2}/2
\end{equation}
The excess delay due the gravitational bending is then $\delta t=\delta
L/c=L\alpha^{2}/2c$. Hence, this effect is of second order in $\alpha$. A more
accurate expression from the lensing equation is $\delta t\simeq
(1+z)\alpha^{2}D_{L}D_{S}/2cD_{LS}$, which corrects for the redshift distance
of the source. The time delay due to the bending of waves propagating at the
velocity $c$ in the gravitational field of a mass distribution of size smaller
than the distance between the source and the observer can now be estimated
referring to the figure 1.

Conservatively taking $3\times10^{14}M_{\odot}$ as the total mass of the Virgo
cluster of galaxies and its dark matter halo within a radius of about 3 Mpc
(and the `impact parameter' of 10 Mpc), the bending angle $\alpha$ is
approximately $6\times10^{-6}$ rad \cite{Unni-DCC}. This is consistent with
the gravitational field of the Virgo cluster at the local group, estimated
from the infall velocity of about 200 km/s of the local group towards Virgo
cluster. The deflection of the path translates to the gravitational bending
delay $L\alpha^{2}/2c>3\times10^{4}\,s,$ where we have taken $L$ as half the
distance to NGC 4993 (this is consistent with the combination of distances
appearing in the lensing equation). This has to be compared to the 1.7 s delay
between the GW from the BNS merger and the GRB. Since the gravitational waves
are pure gravitational energy in propagation, this observed universality of
`free fall' under the gravitational action of the mighty cluster of matter on
both electromagnetic waves and gravitational waves readily proves the Strong
Equivalence Principle, with the tightness of the constraint on any violation
smaller than $\Delta<6\times10^{-5}$. \ As expected, the Shapiro delay
constraint on the SEP that we obtained in this work is by far the most
stringent, $\Delta\leq6\times10^{-10}$.

\section{Concluding remarks}

A unique test of the strong equivalence principle, which asserts the universal
gravitational coupling to the gravitational energy itself, is devised by
recognizing that the propagating gravitational waves are entirely
gravitational energy. A comparison of the propagation of gravitational energy
with gamma photons, from the merger event of the binary neutron stars detected
by the LIGO-Virgo detectors, yielded the most stringent test of the SEP, with
the constraint on any violation $\Delta\leq6\times10^{-10}$. Besides, this is
the only test of the SEP in the radiation sector. The design sensitivity of
the upgraded advanced GW detectors is 170- 300 Mpc for binary neutron star
mergers, which is 3 times the sensitivity they had when the event GW170817
happened. This means that one can expect an order of magnitude higher event
rate at the full sensitivity, expected by 2025, and about 3 to 10 BNS
events/year, with source identification. With many such events, the
statistical precision and confidence in our unique test of the SEP will
steadily improve. In such a scenario, there is no doubt that this
gravitational wave test will remain the most precise confirmation of the
strong equivalence principle.

\end{document}